\documentstyle[preprint,aps,tighten,epsfig]{revtex}

\begin{document}
\draft
\title{The role of electron-screening deformations in solar nuclear fusion
reactions and the solar neutrino puzzle}
\author{Theodore E. Liolios $^{1,2,3}$ \footnote{theoliol@physics.auth.gr}}
\address{$^1$European Center for Theoretical Studies in Nuclear Physics and Related Areas\\
Villa Tambosi, I-38050 Villazzano (Trento), Italy\\
\footnote{Permanent address}$^2$University of Thessaloniki, Department of Theoretical Physics\\
Thessaloniki 54006, Greece\\
$^3$ Hellenic War College, BST 903, Greece\\
}

\date{July 2000}
\maketitle

\begin{abstract}
Thermonuclear fusion reaction rates in the solar plasma are enhanced by the
presence of the electron cloud that screens fusing nuclei. The present work
studies the influence of electron screening deformations on solar reaction
rates in the framework of the Debye-H\"{u}ckel model. These electron-ion
cloud deformations, assumed here to be static and axially symmetric, are
shown to be able to considerably influence the solar neutrino fluxes of the $%
pp$ and the $CNO$ chains, with reasonalble changes in the macroscopic
parameters of the standard solar model (SSM) . Various known deformation
sources are discussed but none of them is found strong enough to have a
significant impact on the SSM neutrino fluxes.
\end{abstract}


\section{Introduction}

As the quest for the solution to the solar neutrino puzzle continues, all
the parameters associated with the production and detection of solar
neutrinos are exhaustively investigated. Naturally, stellar thermonuclear
reactions have attracted a lot of attention in the nuclear physics community
since not only do they govern the evolutionary stages of a star but also
they can possibly hide the solution to the notorious discrepancy between
theoretically and experimentally calculated solar neutrino fluxes.

The effects of the electron-ion screening on stellar fusion cross sections
have drawn a lot of attention \cite{salpeter54}\cite{graboske73}\cite
{mitler77}\cite{dzitko95}, especially in relation to the solar neutrino
puzzle. Various screening prescriptions have been suggested, each of which
has some inherent inadequacies. Salpeter's\cite{salpeter54} weak screening
formula can be safely used in the study of $pp$ solar fusion reactions where
the weak screening regime is obeyed, though a recent study showed\cite
{gruzinov98} that, for most practical purposes, it can reasonably be applied
to other solar reactions as well. On the other hand, the formula given by
Mitler\cite{mitler77} is considered the most reliable as it describes fairly
well all the screening regimes, weak screening ($WES$) , intermediate ($IS$)
and strong screening ($SS$), though the assumption of a constant electron
density around the fusing nuclei is rather arbitrary.

As regards the solar neutrino puzzle, the prevalent belief is that, under
the present standard solar model (SSM), the screening effect is under
control. However, various non-standard solar models have been proposed, in
an attempt to explain the discrepancy between the theoretical and
experimental neutrino fluxes. Some of them are plausible, while others are
exaggerated. Among those attempts, the existence of a primordial magnetic
field in the solar interior has received relatively little attention, while
its relation to the screened fusion reactions themselves has never been
studied.

This lack of interest stems from the disheartening fact that if a large
magnetic field (of the order of $\sim 10^{9}$ Gauss) is added to the
equation of solar structure, the predicted event rate of the Cl$^{37}$
experiment is increased by a factor of two\cite{bahcall71}. As the magnetic
field has the opposite effect from the one that is desired, no further
investigation has been made by the same author\cite{bahcallbook}.
Admittedly, some other studies appeared\cite{bartenwerfer73} which showed
that a combination of differential rotation and strong magnetic fields in
the sun can actually reduce the $Cl^{37\text{ }}$ signal but none of them
has been adapted as a component of the SSM. Nevertheless, as it has been
recently shown , a superstrong magnetic field can accelerate hydrogen fusion
reactions in stars and in the laboratory, an effect which could influence
the solar neutrino fluxes.

This is due to the fact that strong magnetic fields ($B>10^{9}G)$ compress
hydrogen atoms both perpendicular and parallel to the field direction. The
magnetic field ties the electrons to the field lines so that their response
to a Coulomb attraction is essentially restricted to a one-dimensional
motion parallel to the field. It is therefore plausible to assume that the
presence of such a field would modify the screening effect of the electron
cloud just as it happens on the surface of neutron stars. Such screening
deformations can be studied qualitatively by means of the Debye-H\"{u}ckel
model. Actually, in the presence of a large magnetic field the
Debye--H\"{u}ckel radius which represents a spherical distribution of the
electron cloud around the nuclei will become orientation dependent. This
effects causes, inevitably, reaction rates and neutrino fluxes to be
orientation dependent themselves.

The purpose of this paper is to investigate the dependence of thermonuclear
fusion reaction rates on the deformations of the ionic-electron cloud that
screens the reacting nuclei and its possible consequences on solar neutrino
fluxes. The layout of the paper is as follows: In \S II the advantages and
disadvantages of the Debye-H\"{u}ckel potential in the study of solar
nuclear reactions are briefly investigated. In \S III the formalism for an
axially deformed screening cloud is established, underlining its effects on
the $pp$ reaction rates and the associated $pp$ neutrino fluxes. In \S IV
there is given a measure of the uncertainty of the solar neutrino fluxes due
to the presence of such screening deformations. Finally \S V investigates
some potential sources of deformation, while the main results of the present
paper are summarized in \S VI.

\section{Advantages and disadvantages of the Debye-H\"{u}ckel potential}

In the stellar plasma gravitational compression and quantum mechanical
tunneling combine in order to achieve the classically impossible fusion
between light nuclei. The electron gas that surrounds the nuclei acts as a
catalyst in the reaction, by lowering the repulsive Coulomb barrier which
prevents atomic nuclei from approaching each other.

In the framework of the Debye-H\"{u}ckel model each nuclei is assumed to
polarize its neighborhood creating a spherically symmetric but
inhomogeneously charged ionic cloud around it. In this model, the potential $%
V\left( r\right) $ of a given nucleus of charge $Z_{1}$ can be found by the
equation of Poisson : 
\begin{equation}
\nabla ^{2}V\left( r\right) =r_{D}^{-2}V\left( r\right)  \label{poissonrd}
\end{equation}
with 
\begin{equation}
r_{D}^{-2}=\frac{4\pi e^{2}}{kT}\left( \sum_{i}Z_{i}^{2}n_{i}+n_{e}\theta
_{e}\right)  \label{rd}
\end{equation}
where $r_{D}$ is the Debye-H\"{u}ckel radius, $\theta _{e}$ is the electron
degeneracy factor, and $n$ the number densities of ions $\left( n_{i}\right) 
$ with atomic number $Z_{i}$ and electrons $\left( n_{e}\right) ,$
respectively. The solution of Eq. $\left( \ref{poissonrd}\right) $ is a
Yukawa potential which acts in the vicinity of the ion $Z_{1}e$ and has the
form:

\begin{equation}
V_{D}\left( r\right) =\frac{Z_{1}e}{r}\exp \left( -\frac{r}{r_{D}}\right)
\label{vdh}
\end{equation}
A measure of the stellar plasma response to Coulomb interactions is the
plasma coupling parameter, which is defined as: 
\begin{equation}
\Gamma _{ij}=\frac{Z_{i}Z_{j}e^{2}}{akT}  \label{gamma}
\end{equation}
where $a$ is the mean interionic distance, $Z_{ij\text{ }}$the atomic
numbers of the reactants and $kT$ the thermal kinetic energy. Although the
domains of the weak screening $\left( WES\right) $, the intermediate
screening $\left( IMS\right) $ and strong screening $\left( SS\right) $ are
not precisely defined , in an ion fluid one can define them as follows:

\begin{equation}
WES:\Gamma _{ij}<<1,\qquad IMS:\Gamma _{ij}\sim 1,\qquad SS:\Gamma _{ij}>>1
\label{screendef}
\end{equation}
It is now easy to show that the assumption $\Gamma _{ij}<<1$ of the $\left(
WES\right) $ regime for the proton-atom reaction channel between the test
nuclei $Z_{2}e$ and the generators of the above potential, i.e. nuclei $%
Z_{1}e$ plus ionic cloud, can be written:

\begin{equation}
\frac{Z_{1}Z_{2}e^{2}}{r_{D}kT}<<1  \label{wescond}
\end{equation}
The above condition has generated a lot of controversy in the study of solar
fusion cross sections as it is violated for solar nuclear reactions with $%
Z_{1}Z_{2}$ larger than unity.

Another interesting fact is that according to the above model for small $r$
the ion density now vanishes while the electron density diverges. By noting
this, Mitler\cite{mitler77} assumed the above model to be valid only beyond
some radius $r_{1}$ while for shorter distances he assumed the that the
electron density around the nucleus is constant and equal to the mean
electron density in the plasma, $n_{e}\left( \infty \right) $.

However this is also an oversimplification, especially for the heavier
nuclei of astrophysical interest where it underestimates the electron charge
density around the ion. For instance near a $Be^{7}$ nucleus in central
solar conditions we have a density , roughly $3.8en_{e}\left( \infty \right) 
$.

Moreover, the model in question is forced to undergo another compromise when
calculating the thermalized cross section $\left\langle \sigma
v\right\rangle ^{sc}$which appears in the screened reaction rate between
nuclei $i$ and $j$ :

\begin{equation}
r_{ij}^{sc}=\left( 1+\delta _{ij}\right) ^{-1}n_{i}n_{j}\left\langle \sigma
v\right\rangle ^{sc}  \label{rijdelta}
\end{equation}
where $n_{i},n_{j}$ are the number densities of nuclei $i$ and $j\,\,$%
respectively, and $\delta _{ij}$ is the Kronecker delta. As the WKB integral
involved cannot be found analytically, we inevitably resort to a linear
approximation of potential $\left( \ref{vdh}\right) $ . However even if we
don't resort to that approximation and take into account non-linearities\cite
{liolios00} of the Debye-H\"{u}ckel potential this would add a negligible
contribution to Salpeter's $\left( WES\right) $ prescription. Note that
whenever the $\left( WES\right) $ condition is challenged one should always
investigate non-linear corrections as will be shown in the study that
follows.

Finally, the spherical symmetry around the two reactants is taken for
granted by the standard Debye-Huckel model. The fact that no deformations
are assumed for the ionic-electron cloud around the point like nucleus can
obviously be the source of uncertainties in the calculation of stellar
reaction rates. It is the principle objective of this paper to investigate
the effects of such deformations on reaction rates and the associated solar
neutrino fluxes, without focusing in detail on the sources of deformations
themselves.

\section{Electron screening deformations}

In the adiabatic model the target and the projectile nuclei are assumed to
be surrounded by a static, spherical electron cloud, whose electron charge
density falls off exponentially with respect to the distance from the center
of the cloud which is the nucleus itself. In central solar conditions the
mean ion velocity $\left\langle u_{i}\right\rangle =\left( 8kT/\pi \mu
\right) ^{1/2}$ is roughly\cite{dzitko95} fifty times smaller than the mean
electron velocity $p_{e}^{2}/2m_{e}\simeq \left( 3/2\right) kT$ thus
justifying the fact that as the nucleus moves the electron cloud has enough
time to re-arrange itself so that it practically screens the nucleus at all
times. However oscillations of the ionic cloud are inevitable due to their
speed which is much lower than that of the electrons. For the main sequence
stars Mitler\cite{mitler77} showed that in the framework of the standard
solar model (SSM) the distortion of the common charge clouds has only a
small effect on the screening calculations $\left( \sim 2\%\right) .$

Moreover, the possible presence of a strong magnetic field is bound to cause
substantial deformations of the electron cloud by compressing it both
parallel and perpendicular to the field direction. On the other hand the
existence of other, as yet unspecified, sources of deformations cannot be
ruled out.

Therefore treating the screening cloud as a rigid (albeit inhomogeneous)
sphere is an assumption which must by further investigated, especially when
primordial magnetic fields are considered.

Studies of heavy nuclei fusion reactions have shown that theoretical
predictions of cross section can be greatly improved\cite{moeller94} by
assuming rotations and deformations of the fusing nuclei. It is therefore
plausible to consider similar effects in the study of screened thermonuclear
reactions where the electron cloud is assumed to be deformed. In fact this
deformation can be parametrized in the framework of the liquid-drop model so
that the Debye-H\"{u}ckel radius is considered a measure of the electron
cloud. The deformed DH radius is now:

\begin{equation}
r_{D}\left( \theta ,\phi \right) =r_{D}^{\left( 0\right) }\left[
1+\sum_{m}\beta _{m}Y_{2}^{m}\left( \theta ,\phi \right) \right]
\label{rdeformed}
\end{equation}
where $r_{D}^{\left( 0\right) },$ takes care of volume conservation and $%
Y_{2}^{m}$ is the usual spherical harmonic function. For simplicity and
reasons that will soon become clear only quadrupole deformations will be
considered. Moreover, we assume that the deformation is axially symmetric
and take the $z$ axis along the axis of symmetry. Disregarding the
rotational degree of freedom we obtain the surface shape

\begin{equation}
r_{D}\left( \theta \right) =r_{D}^{\left( 0\right) }\left[ 1+\beta
Y_{2}^{0}\left( \cos \theta \right) \right]  \label{rdtheta}
\end{equation}
where the angle $\theta $ is measured from the axis of symmetry i.e. the $z$
axis. Note that $r_{D}^{\left( 0\right) }\left( \beta \right) \simeq
r_{D}^{\left( 0\right) }\left( -\beta \right) .$

For $\beta >0$ the single axis is larger than the double axis and the cloud
is a prolate spheroid, that is cigar-shaped. For $\beta <0$ the single axis
is smaller than the double axis and the cloud is an oblate spheroid i.e.
disk-shaped.

Note that the weak screening approximation restricts the possible values of $%
\beta .$ A reasonable assumption which stems from Eq.$\left( \ref{wescond}%
\right) $ is that in the $WES$ regime the following relation must be
fulfilled:

\begin{equation}
\frac{Z_{1}Z_{2}e^{2}}{r_{D}\left( \theta \right) kT}\leq 0.1
\end{equation}
In solar conditions for the $pp$ reaction where the use of the $WES$
formalism is incontrovertible we have

\begin{equation}
\frac{e^{2}}{r_{D}kT}\simeq 0.05  \label{scterm}
\end{equation}
Therefore , for all orientations, the following inequality must hold :

\begin{equation}
-0.8\leq \beta \leq 0.8
\end{equation}
where we have disregarded the contribution of volume conservation which is
always less than $5\%$. The deformation parameter can now take all the above
values without violating the $WES$ condition $\left( \ref{wescond}\right) $
, thus rendering the use of the deformed screening formalism legitimate.

If we take into account the nuclear potential $V_{N}\left( r\right) $ then
the total potential of the reaction is

\begin{equation}
V\left( r;\theta ;\beta \right) =V_{N}\left( r\right) +V_{D}\left( r;\theta
;\beta \right) +\frac{\hbar ^{2}}{2\mu r^{2}}l\left( l+1\right)
\end{equation}
where the centrifugal term is assumed to be independent of the orientation.
This assumption is immaterial here as we will only consider very low-energy
reactions where $s$-interactions dominate. In that case the orientation
dependent cross section of the nuclear fusion reaction is given by:

\begin{equation}
\sigma \left( E;\theta ;\beta \right) =\frac{S\left( E\right) }{E}P\left(
E;\theta ;\beta \right)  \label{cstheta}
\end{equation}
where

\begin{equation}
P\left( E;\theta ;\beta \right) =\exp \left[ -\frac{2\sqrt{2\mu }}{\hbar }%
\int_{R}^{r_{c}\left( \theta ;\beta \right) }\sqrt{V_{D}\left( r;\theta
;\beta \right) -E}dr\right]  \label{penfact}
\end{equation}
and $r_{c}\left( \theta ;\beta \right) $ is the classical turning point
given by

\begin{equation}
V_{D}\left( r_{c};\theta ;\beta \right) =E
\end{equation}
The thermonuclear reaction which can be studied safely by means of the above
formalism is the one that dominates the solar neutrino production namely: $%
H^{1}\left( p,e^{+}\nu _{e}\right) H^{2}.$ For that reaction, in the
undeformed weak screening case, it turns out that in the region of the
maximum energy production $R=0.09R_{\odot \text{ }}$the Gamow peak is $%
E_{0}=5.599\,keV$ , the classical turning point is roughly $r_{c}=0.01r_{D%
\text{ , }}$while the $WES$ enhancement factor is to good approximation: $%
f_{0}^{WES}=1.049.$

Actually, non-linear screening corrections\cite{liolios00} can be safely
neglected. Along the whole profile of orientations and $\beta $ parameters
the contribution of higher order terms to the shifts of the screening
corrections and the Gamow peak has been found negligible. Hence, screening
deformation effects can be simply represented by Salpeter's\cite{salpeter54} 
$WES$ formula

\begin{equation}
f_{D}\left( \theta ;\beta \right) =\exp \left( \frac{e^{2}}{kTr_{D}\left(
\theta \right) }\right)
\end{equation}
where the DH radius $r_{D}\left( \theta \right) $ is now orientation
dependent. Finally we obtain:

\begin{equation}
f_{D}\left( \theta ;\beta \right) =\left( f_{0}^{wes}\right) ^{g^{-1}\left(
\theta ;\beta ,\right) }
\end{equation}
where $g\left( \theta ;\beta \right) $ is the ratio $r_{D}\left( \theta
;\beta \right) /r_{D}$:

\begin{equation}
g\left( \theta ;\beta \right) =\left\{ \frac{1}{2}\int_{-1}^{1}\left[
1+\beta \sqrt{\frac{5}{16\pi }}\left( 3u^{2}-1\right) \right] ^{3}du\right\}
^{-1/3}\left[ 1+\beta \sqrt{\frac{5}{16\pi }}\left( 3\cos ^{2}\theta
-1\right) \right]
\end{equation}

In Fig.1 the orientation dependent DH radius $r_{D}\left( \theta ;\beta
\right) $, measured in units of $r_{D},$ is plotted in polar coordinates
with respect to the azimuthal angle $\theta $ and the deformation parameter $%
\beta $ . It is obvious that along the axis of symmetry of the cloud, $z$%
-axis $\left( \theta =0\right) ,$ a positive $\beta \,$parameter ''stretches
out'' the ionic cloud (a prolate spheroid shape) while a negative $\beta $
parameter ''sucks in'' the cloud (an oblate spheroid shape). As one would
expect, for both positive and negative parameters the larger the absolute
value of $\beta $ the more pronounced the deformation.

Note that Fig. 1 actually represents the deformation factor $g\left( \theta
;\beta \right) $ while the corresponding classical turning point is $%
r_{c}\left( \theta ;\beta \right) =261g\left( \theta ;\beta \right) fm.$ For
example for a $\beta =0.8$ deformation a proton cruising along the z-axis in
the plasma with an energy equal to the Gamow peak $E_{0}=5.56$ $keV$ will
come up against the potential wall at a distance roughly $1.43$ times
further than in the undeformed case $\left( r_{c}=261fm\right) $, that is $%
r_{c}\left( \theta =0,\beta =0.8\right) =373fm$. On the other hand for a $%
\beta =-0.8$ the classical turning point is reduced to $r_{c}\left( \theta
=0,\beta =-0.8\right) =123fm.$

The screened Coulomb potential $V\left( r;\theta ;\beta \right) $ can be
visualized by means of Fig. 2 where we have plotted in polar coordinates the
deformed shapes of the potential at a distance $r=r_{D}$ from the origin. At
that distance the potential contours $V_{D}\left( r_{D};\theta ;\beta
\right) $ of Fig. 2 are only a function of the orientation. Hence, as a
proton enters the ionic cloud of the Hydrogen atom on its way to fusion,
according to the angle at which it enters the cloud it will experience a
different (thicker or thinner) potential wall . The thicker the wall, the
most improbable the reaction and of course the smaller the reaction rate.

The orientation dependent acceleration of the reaction is displayed in Fig.
3 where the screening factor is plotted with respect to the azimuthal angle
and the deformation parameter for the $pp$ reaction at $R=0.09R_{\odot }$.
The reaction rate can be $1.1$ times faster if the proton enters a
disk-shaped cloud $\left( \beta =-0.8\right) $ at zero angle. On the other
hand a much slower reaction is obtained for a cigar-shaped ionic cloud
(almost no enhancement at all for $\beta =0.8)$.

The impact of screening deformation on the neutrino fluxes will be discussed
in a more quantitative way in the section that follows.

\section{Solar neutrino fluxes.}

In the solar region of maximum energy production $\left( R/R_{\odot
}=0.09\right) $ the thermal kinetic energy is $kT=1.161keV$ while for the $%
pp $ reaction the standard (undeformed) weak-screening factor is $%
f_{0}^{WES}=1.049.$ On the other hand for a $\beta =-0.4$ deformation and an
angle of impact of $\theta =0$ the screening factor is $f_{D}=1.067.$ This
corresponds to an acceleration of the (undeformed) weakly screened $pp$
reaction by roughly $1.7\%\,$which in turn reflects on the cross-section
factor given by $S_{D}=f_{D}S.$ As the principal source of energy in the Sun
is the $pp$ reaction this acceleration would influence both the solar
structure and the neutrino fluxes by reducing the central temperature and
density in order to conserve luminosity. (An account of what happens in the
sun if the cross-section factor $S_{pp}$ increases can be found in Ref. .%
\cite{castellani93})$.$

In most solar evolution codes the $pp$ screening factor is evaluated by
means of Salpeter's formula which has been proved to be valid and accurate
in standard conditions. In the deformed case the quantity $f_{D}$ should be
used, instead. We can obtain an approximation of the uncertainties
introduced due to the presence of such deformations by using the
proportionality formulae\cite{bahcallbook}\cite{ricci95} which relate
neutrino fluxes to screening factors. In order to isolate the $pp$
uncertainty, we will assume that except for the $pp$ reaction all the other
neutrino-producing reactions remain unaffected by the deformations, thus
obtaining a minimum of the total associated uncertainty.

For various solar fusion reactions the ratios of the deformed neutrino
fluxes $\Phi ^{D}$ to the ones obtained in the $WES$ regime $\Phi ^{WES},$
are as follows:

$H^{1}\left( p,e^{+}\nu _{e}\right) H^{2}$

\begin{equation}
\left( \frac{\Phi _{pp}^{D}}{\Phi _{pp}^{WES}}\right) _{pp}=\left( \frac{%
f_{D}}{f_{0}^{WES}}\right) ^{0.14}  \label{fpp}
\end{equation}

$H^{1}\left( pe^{-},\nu _{e}\right) H^{2}$

\begin{equation}
\left( \frac{\Phi _{hep}^{D}}{\Phi _{hep}^{WES}}\right) _{pp}=\left( \frac{%
f_{D}}{f_{0}^{WES}}\right) ^{-0.08}
\end{equation}

$Be^{7}\left( e^{-},\nu _{e}\right) Li^{7}:$

\begin{equation}
\left( \frac{\Phi _{Be^{7}}^{D}}{\Phi _{Be^{7}}^{WES}}\right) _{pp}=\left( 
\frac{f_{D}}{f_{0}^{WES}}\right) ^{-0.97}  \label{fbe}
\end{equation}

$Be^{7}\left( p,\gamma \right) B^{8}\left( e^{+},\nu _{e}\right) B^{8*}:$

\begin{equation}
\frac{\Phi _{B}^{D}}{\Phi _{B}^{WES}}=\left( \frac{f_{D}}{f_{0}}\right)
^{-2.6}
\end{equation}

$N^{13}\left( e^{+}\nu _{e}\right) C^{13}$ and $O^{15}\left( e^{+},\nu
_{e}\right) N^{15}:$

\begin{equation}
\frac{\Phi _{N,O}^{D}}{\Phi _{N,O}^{WES}}=\left( \frac{f_{D}}{f_{0}^{WES}}%
\right) ^{-22/8}  \label{fno}
\end{equation}
According to the above formulae, the presence of a screening deformation of $%
\beta =-0.4$ and an angle of impact of $\theta =0$ can perturb the solar
neutrino fluxes calculated in the SSM by at least $4.5\%$ for the $N,O$ and $%
B^{8}$ neutrinos and by less than $1\%$ for the less sensitive $pp,hep,$and $%
B^{7}$neutrinos. Such a source of deformation would also perturb the
electron cloud around the heavier nuclei involved in neutrino production,
thus modifying their screening factors which we arbitrarily considered
constant here.

It is very tempting to investigate what happens if the deformations are
stronger. For a collision along the z-axis with $\beta =-0.8$ we obtain $%
f_{D}=1.16$ and the corresponding uncertainties are at least $23\%$ for the $%
N,O$ and $B^{8}$ neutrinos, $9.2\%$ for the $Be^{7}$ neutrinos and less than 
$2\%$ for the $pp,hep$ ones. Adding the effects of the screening factors of
the other reactions, which have been disregarded so far, the uncertainties
can be dramatic. It seems therefore that the presence of screening
deformations in the sun can tune the predicted neutrino fluxes in order to
reduce the observed deficit.

It is crucial to study what deviations from the SSM parameters such
deformations can induce. As was previously noted conservation of luminosity
implies a reduction in the central temperature $T_{c}$ as a result of an
increase in $S_{pp},$ since $L_{\odot }\sim S_{pp}T_{c}^{8}$. This implies:

\[
\frac{T_{c}^{D}}{T_{c}^{WES}}=\left( \frac{f_{D}}{f_{0}^{WES}}\right) ^{-%
\frac{1}{8}} 
\]

Therefore, for the weaker (stronger) of the above considered deformations
the $\left( WES\right) $ central temperature of the sun would have to
decrease by $0.2\%$ $\left( 1.2\%\right) $. On the other hand, as the ratio $%
\rho /T^{3}$ is approximately constant along the solar profile, the new
central density would now be given by:

\[
\frac{\rho _{c}^{D}}{\rho _{c}^{WES}}=\left( \frac{f_{D}}{f_{0}^{WES}}%
\right) ^{-\frac{3}{8}} 
\]
that is roughly decreased by $0.6\%\,\,\left( 3.7\%\right) $ with respect to
the $\left( WES\right) $ assumption$.$ Both values represent reasonable
deviations from the $\left( WES\right) $ SSM considering that the $\left(
WES\right) $ assumption itself causes a $0.6\%$ deviation from the $\left(
NOS\right) $ central temperature and another $1.8\%$ deviation from the $%
\left( NOS\right) $ central density.

\section{Screening deformation sources}

Having established the general formalism for the investigation of screening
deformations, a discussion of potential sources of such deformations is
imperative. In a study of the effects of superstrong magnetic fields in the
stellar plasma there was shown clearly that even at zero energies the
screening electron cloud is deformed , in the sense that it becomes
compressed perpendicular and parallel to the field direction. This
deformation can dramatically accelerate hydrogen fusion reactions not only
in neutrons stars\cite{heyl96} but also in the laboratory as was shown
recently\cite{liolios2000},\cite{lioliosepj}. It is very plausible therefore
to assume that the presence of a superstrong magnetic field in the solar
interior would cause similar effects.

From the present work it is obvious that if a substantial correction to the
solar neutrino fluxes is to be made by means of screening deformations those
have to be larger than $\beta =-4$. To gain an idea of what kind of solar
magnetic field would cause such deformation in the center of the sun we can
use the results of Ref. \cite{lioliosepj} where it shown clearly that any
magnetic field weaker than the Intense Magnetic Field Regime:

\[
B_{IMF}=4.7\times 10^{9}G 
\]
would have no effect on hydrogen fusion reactions. Hence, in order to
decrease the solar neutrino fluxes by means of magnetically catalyzed
thermonuclear fusion, the magnetic field required must be stronger than $%
10^{10}G.\,$

Admittedly such a superstrong field cannot be easily justified. After the
disheartening result\cite{bahcall71} that the presence of a strong magnetic $%
\left( \sim 10^{9}G\right) $ in the solar interior increases the predicted
neutrino fluxes (doubles the $Cl^{37}$ signal by increasing the pressure
gradient in the sun) few investigators have looked into the matter. This is
also due to some additional arguments which indicate the implausibility of a
solar magnetic field larger than $10^{9}G.$ Such arguments include the
limiting strength set by Chandrasekhar and Fermi\cite{fermi53}, stability
reasons\cite{cowling65} and magnetic buoyancy\cite{parker74}, though there
is a very interesting work\cite{bartenwerfer73} which argues that a
combination of a differential rotation and magnetic field can reduce the $%
Cl^{37\text{ }}$signal opposing the results of Ref. \cite{bahcall71}. Note
that Helioseismology is another concern when considering such a strong
magnetic field in the sun. Nevertheless, such a field, despite stability and
buoyancy counter-arguments, can have been formed by the interstellar
magnetic field which was frozen into the matter out of which the sun was
formed, or there may be an unspecified mechanism of continuous generation.

It is now obvious that, regarding the solar neutrino puzzle, even if we
accept the presence of a superstrong magnetic field the corrections induced
are much smaller than the ones required to reconcile theory and experiment.
Therefore, taking also into account other counter-arguments, it seems that
magnetically induced screening deformations (magnetic catalysis) cannot
possibly be the answer to the neutrino puzzle.

Another plausible source of screening deformations is the fact that in the
solar center the average interionic spacing is $a\simeq 2.8\times 10^{-9}cm,$%
which is similar to the Debye radius. Therefore the spherical electron cloud
assumption is not well justified. The cloud is more likely to assume an
ellipsoidal distribution around the two reactants, like a fissioning nucleus 
\cite{dzitko95}. However, the argument that the incoming fusing nucleus will
be carrying its own cloud, suggested in Ref. \cite{shaviv}, and Ref. \cite
{dzitko95}, thus increasing the deformations, doesn't seem to have a
significant effect as in that case the total of the nucleus-nucleus,
nucleus-cloud, and cloud-cloud interactions would be\cite{shaviv}:

\begin{equation}
V\left( r\right) =\frac{Z_{1}Z_{2}e^{2}}{r}-\frac{3}{2}\frac{Z_{1}Z_{2}e^{2}%
}{r_{D}}
\end{equation}
where the cloud-cloud interaction, in analogy to the recent results for the
laboratory$\cite{lioliosepj},$ would be much smaller than the screening
shift itself. Therefore the screening shift would have been at least $50\%$
larger than the $\left( WES\right) $ one used in standard DH theory thus
accelerating for example the hydrogen fusion reaction by roughly $2.5\%$
with respect to the WES regime. This would result in an increase of the $%
\left( NOS\right) $ $pp$ neutrino fluxes of the order of $1\%$ and a similar
decrease in the $\left( NOS\right) $ central solar temperature. It is
therefore obvious that any screening deformations due to SSM cloud-cloud
interactions in the solar plasma are negligible.

Finally a non-Maxwellian distribution of velocities, such as the flat
Maxwellian\cite{bell}, where the relative particle motion is frozen to $0\,K$
along a specific direction , will establish a preferential direction of
motion for ions thus inducing electron cloud deformations. In fact any kind
of deviation from statistical equilibrium is a source of screening
deformations. For example it has been shown\cite{clayton} that a progressive
depletion of the Maxwellian tail can yield a neutrino counting rate below 1
SNU but a physical cause for that distribution has not been found. Taking
into account other counter-arguments\cite{bahcallbook} to the existence of a
non-Maxwellian distribution in the sun this non-standard source of
deformation has to be deferred to a forthcoming article where the issue will
be studied in detail.

\section{Conclusions}

In this work we investigate the response of the proton-proton reaction to
electron-ion screening deformations in the solar plasma. Those deformations
are studied in the framework of the Debye-H\"{u}ckel model and the results
show that they can induce an orientation-dependent thermalized cross section
which causes the solar neutrino fluxes to be orientation-dependent
themselves. Therefore, in principle, screening deformations can influence
the solar neutrino fluxes with reasonable deviations from the macroscopic
values of the SSM.

Various potential deformation sources are discussed but none of them is
found capable of inducing deformations strong enough to have a significant
impact on the SSM neutrino fluxes. However, the existence of other, as yet
unspecified, deformation sources cannot be ruled out. It seems therefore
necessary to further investigate the possible presence of such sources which
could cause a substantial degree of uncertainty to the solar neutrino fluxes.

Regarding the novelties of the present paper they can be summarized as
follows:

The effects of deformations have received a minimal attention by only two
authors (Ref. \cite{mitler77} and Ref. \cite{dzitko95} ). They both
concluded that the effect is small but none of them studied deformations in
a non-standard solar model as they assumed SSM conditions from the very
beginning of their calculations. The present paper studies screening
deformations in a way completely independent of the source and solar model
conditions. From now on, each time a non-standard source of deformation
appears, it can possibly be connected to the deformation parameter $\beta $
introduced here, just as it happens in heavy ion fusion reactions. However,
that task is admittedly not a trivial one. Moreover, the effects of
magnetically catalyzed fusion on the SSM have been discussed here for the
first time and eventually overruled as a solution to the solar neutrino
puzzle. Finally, here for the first time there is shown that, if
sufficiently strong, screening deformation can actually perturb considerably
the SSM neutrino fluxes with reasonable changes in the central solar
temperature and density. That result warrants further research into the
origin and the effects of such screening deformation sources in the solar
interior.

\section{\protect\medskip {\bf ACKNOWLEDGMENTS}}

This work was financially supported by the Hellenic State Grants Foundation
(IKY) under contract \#135/2000. It was initiated at the Hellenic War
College while its revised version was written at ECT$^{*}$ during a nuclear
physics fellowship. The author would like to thank the director of ECT$^{*}$
Prof. R. Malfliet for his kind hospitality and support.

\newpage {\bf FIGURE CAPTIONS}

Figure 1.

The variation of the orientation dependent DH radius $r_{D}\left( \theta
;\beta \right) $, measured in units of $r_{D},$ with respect to the
azimuthal angle $\theta $ and the deformation parameter $\beta $ in polar
coordinates.

Figure 2.

The orientation dependent DH\ potential $V_{D}\left( r;\theta ;\beta \right)
,$ measured in units of the screening term of Eq.$\left( \ref{scterm}\right) 
$, at a distance $r=r_{D}$ with respect to the azimuthal angle $\theta $ and
the deformation parameter $\beta $ in polar coordinates. The effect is
calculated for the $pp$ reaction in the region of the maximum energy
production $R=0.09R_{\odot }$ where $r_{D}=25719fm$ and $\frac{e^{2}}{r_{D}kT%
}\simeq 0.056$ $keV.$

Figure 3.

The variation of the orientation dependent screening factor $f_{D}\left(
\theta ;\beta \right) $ with respect to the azimuthal angle $\theta $ and
the deformation parameter $\beta $ in polar coordinates. The effect is
calculated for the $pp$ reaction in the region of the maximum energy
production $R=0.09R_{\odot }$ where $r_{D}=25719fm.$

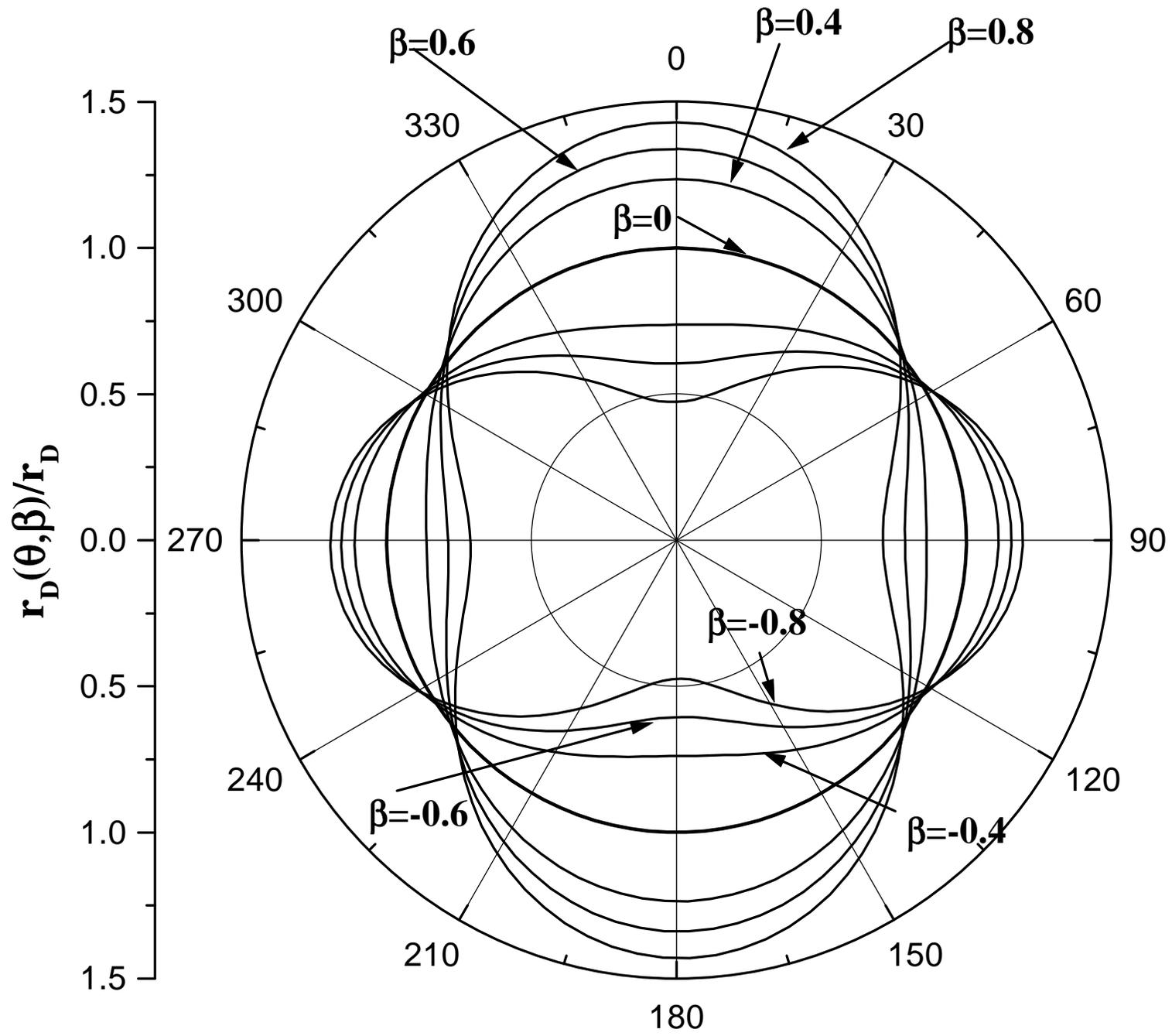
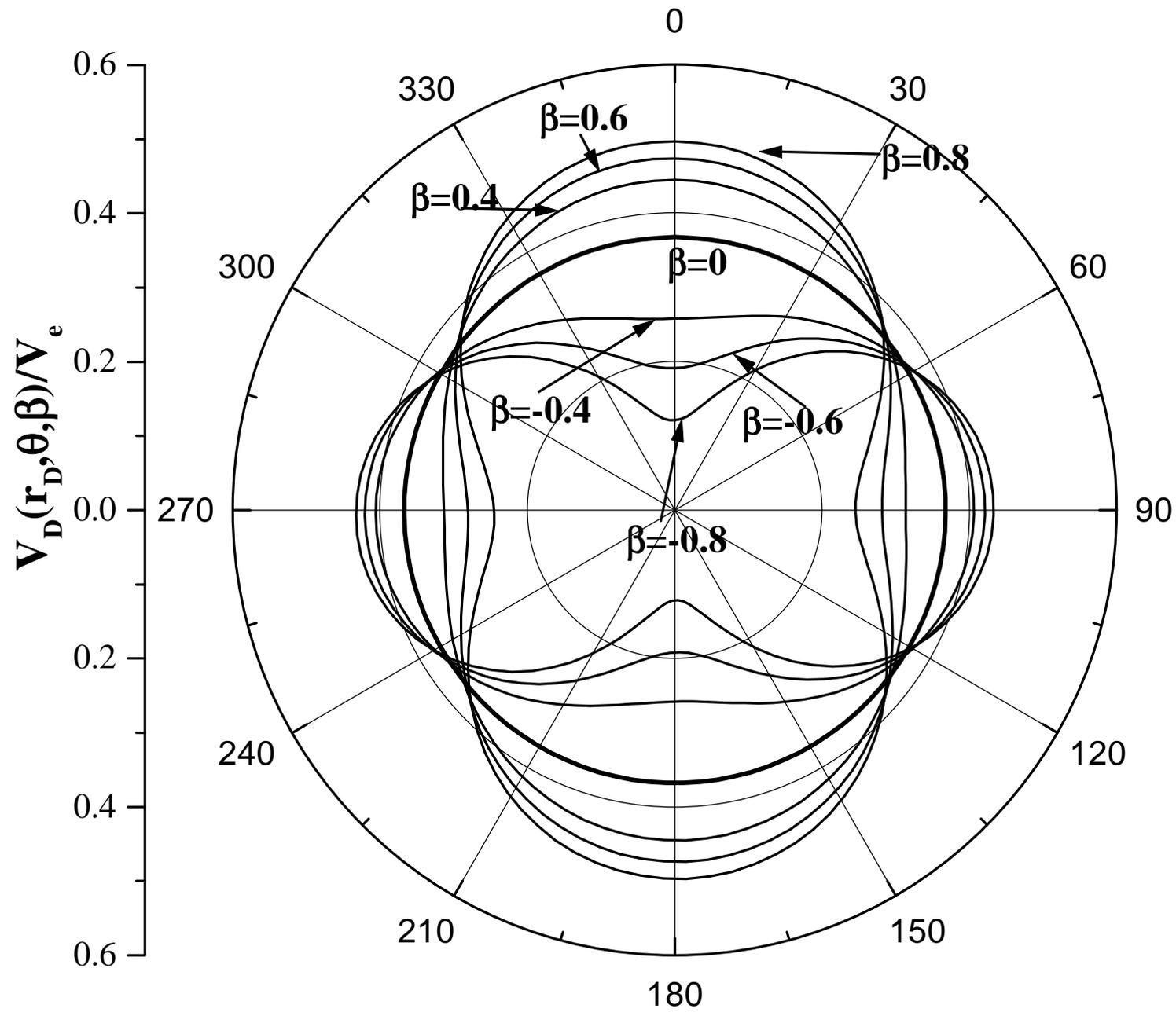
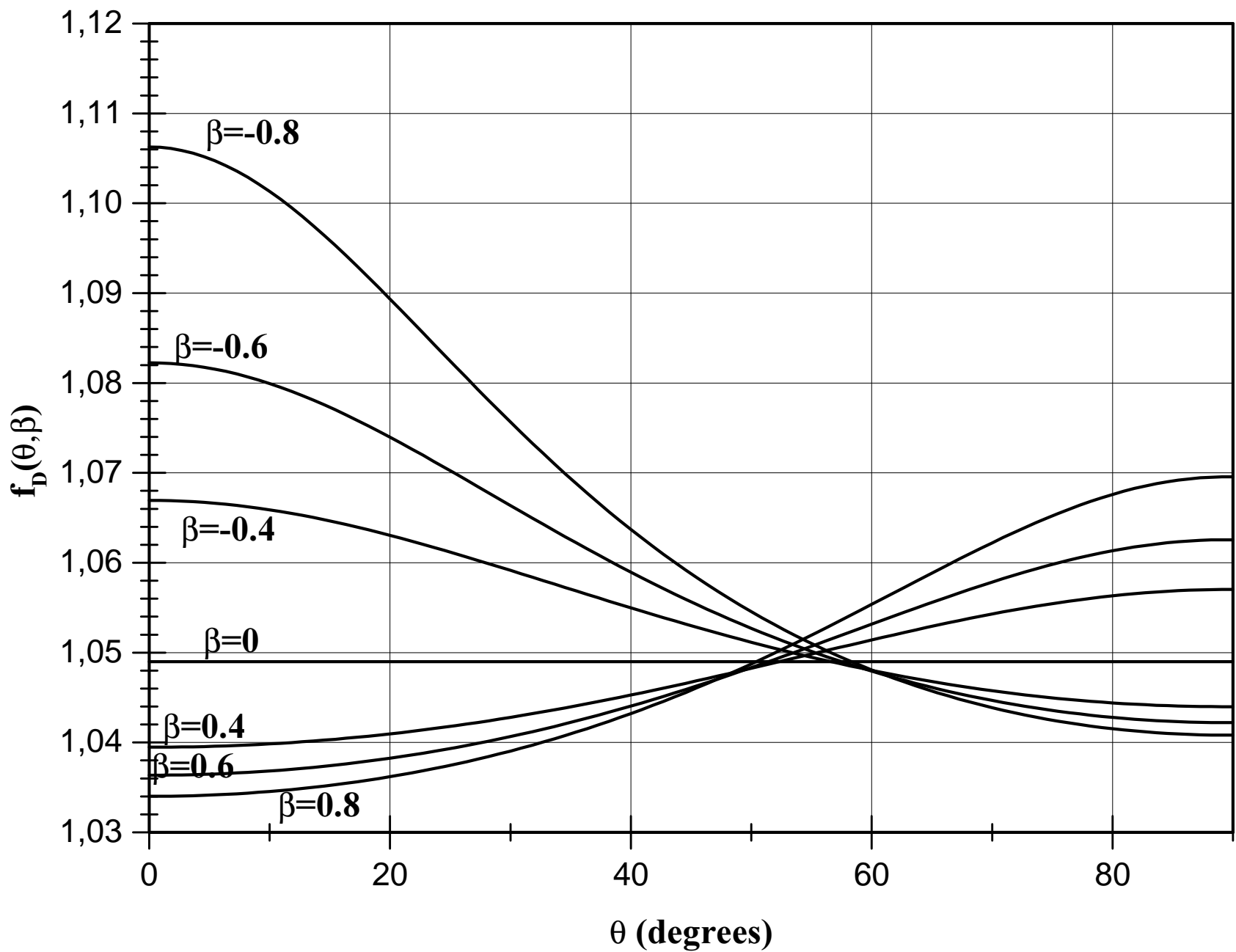
\end{document}